\documentclass[10pt,journal,compsoc]{IEEEtran}

\usepackage{caption}
\usepackage{subcaption}
\usepackage{graphicx}
\usepackage{pgfplots}
\usepackage{booktabs}
\usepackage{multirow}

\ifCLASSOPTIONcompsoc
  \usepackage[nocompress]{cite}
\else
  \usepackage{cite}
\fi
\ifCLASSINFOpdf
\else
\fi
\begin{document}
%

\title{A Survey on Multimodal Wearable Sensor-based Human Action Recognition}

%
%
%
%

\author{Jianyuan~Ni,~\IEEEmembership{Student Member,~IEEE,}
        Hao~Tang,~\IEEEmembership{Senior Member,~IEEE,}
        Syed~Tousiful~Haque,~\IEEEmembership{Student Member,~IEEE,}
        Yan~Yan,~\IEEEmembership{Senior Member,~IEEE}
        and~Anne~H.H.~Ngu,~\IEEEmembership{Member,~IEEE}
\IEEEcompsocitemizethanks{\IEEEcompsocthanksitem J.,Ni and A.H., Ngu are with the Department of Computer Science,Texas State University, San Marcos, TX 78666, USA. \protect\\
E-mail: j\_n317@txstate.edu
\IEEEcompsocthanksitem H.Tang is with the Robotics Institute, Carnegie Mellon University, Pittsburgh, PA 15213, USA.
\IEEEcompsocthanksitem Y. Yan is with the Department of Computer Science, Illinois Institute of Technology, Chicago, IL 60616, USA.
}
\thanks{Manuscript received April, 2024}}

%
%

\markboth{Journal of \LaTeX\ Class Files, 2024}%
{Shell \MakeLowercase{\textit{et al.}}: Bare Demo of IEEEtran.cls for IEEE Journals}
\IEEEtitleabstractindextext{%
\begin{abstract}
The combination of increased life expectancy and falling birth rates is resulting in an aging population. Changes associated with aging, can impact an individual’s quality of life, potentially leading to injuries, mental health issues, or reduced physical activity. Wearable Sensor-based Human Activity Recognition (WSHAR) emerges as a promising assistive technology to the healthy living of older individuals, unlocking vast potential for human-centric applications. However, recent surveys in WSHAR have been limited, focusing either solely on deep learning approaches or on a single sensor modality. In real life, our human interact with the world in a multi-sensory way, where diverse information sources are intricately processed and interpreted to accomplish a complex and unified sensing system. To give machines similar intelligence, multimodal machine learning, which merges data from various sources, has become a popular research area with recent advancements. In this study, we present a comprehensive survey from a novel perspective on how to leverage multimodal learning to WSHAR domain for newcomers and researchers. We begin by presenting the recent sensor modalities as well as deep learning approaches in HAR. Subsequently, we explore the techniques used in present multimodal systems for WSHAR. This includes inter-multimodal systems which utilize sensor modalities from both visual and non-visual systems and  intra-multimodal systems that simply take modalities from non-visual systems. After that, we focus on current multimodal learning approaches that have applied to solve some of the challenges existing in WSHAR. Specifically, we make extra efforts by connecting the existing multimodal literature from other domains, such as computer vision and natural language processing, with current WSHAR area. Finally, we identify the corresponding challenges and potential research direction in current WSHAR area for further improvement.
\end{abstract}

}

\maketitle

\IEEEdisplaynontitleabstractindextext

%
\IEEEpeerreviewmaketitle

\ifCLASSOPTIONcompsoc
\IEEEraisesectionheading{\section{Introduction}\label{sec:introduction}}
\else

\section{Introduction}
\label{sec:introduction}
\fi

%
%
%
%
\IEEEPARstart{T}{he} global population aged 60 or over is expanding at an unprecedented rate. According to the World Population Report, life expectancy at birth is projected to increase from 71 years in 2010–15 to 77 years in 2045–50 \cite{desa2017world}. This demographic shift poses a challenge for most societies, as they strive to ensure their health systems are equipped to adapt. Efforts have been made to maintain or improve the quality of life for older individuals. These include the development of new systems that leverage medical and assistive technologies for long-term care provision, as well as the creation of age-friendly environments. In recent years, the advancement of sensors, wireless communication, and machine learning techniques have spurred the development of assistive technologies. These technologies promote independent, active, and healthy aging \cite{kuerbis2017older}.

\begin{figure*}[t]
  \centering
  \includegraphics[width=\textwidth]{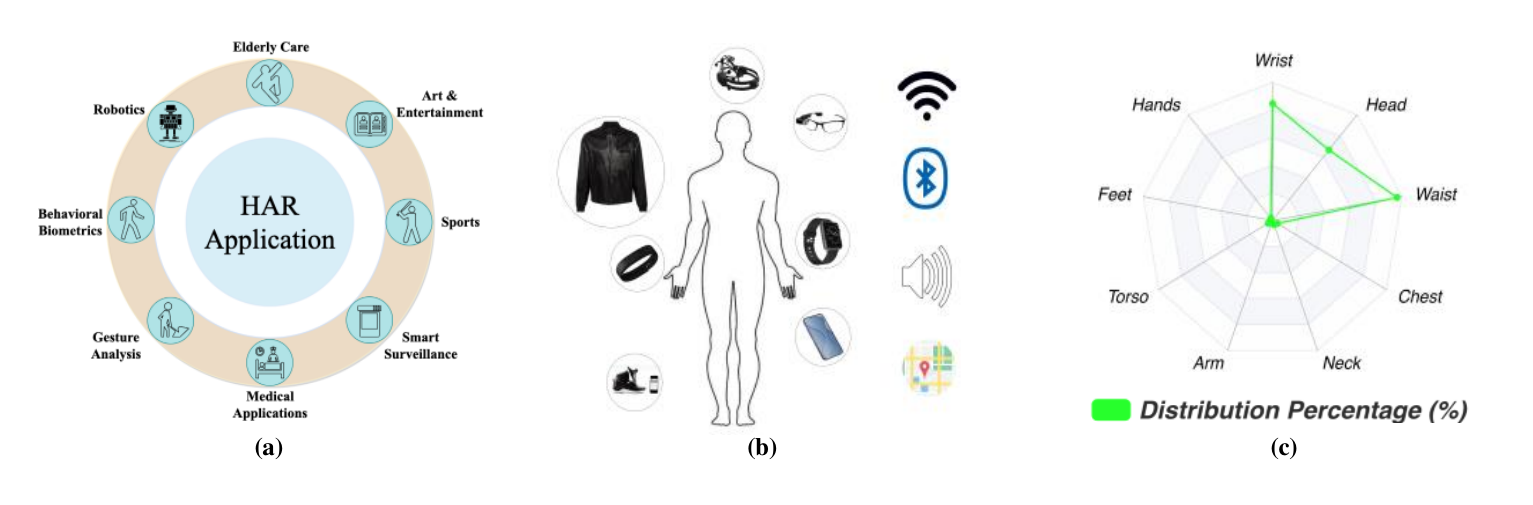}
  \caption{ (a) Applications of wearable sensor-based human activity recognition (HAR). (b) Typical wearable devices for the WSHAR problem. (c) Distribution of wearable devices placed on human body areas \cite{yadav2021review}.}
  \label{figure1}
  \vspace{-0.3 cm}
\end{figure*}

Among these techniques, human action recognition (HAR) is a significant field in extracting deep insights about human behavior from raw sensor data, enabling computing systems to monitor, analyze, and assist individuals in their daily lives. As a result,various HAR systems present in many applications, including video surveillance \cite{liu2021semantics}, human-robot interaction \cite{rodomagoulakis2016multimodal}, and healthcare for the elderly \cite{de2017mobile,maray2023transfer} as shown in Figure \ref{figure1}.a. At present, two primary types of HAR systems are in common use: video-based systems and wearable sensor-based systems. The first type relies on visual modalities such as RGB video, skeleton, and depth data, whereas wearable sensor-based systems use sensors like gyroscope, accelerometer, etc \cite{yadav2021review}. Currently, HAR research is dominated by video-based approaches that contains richer information and can
capture scene context \cite{sun2022human}. While there have been substantial advancements in HAR methods that rely on visual modalities, they have led to increasing privacy concerns due to the utilization of video/image data. For example, purely visual-based systems may not be suitable for areas where privacy is a priority \cite{lyu2017privacy}. Moreover, these video-based systems are unable to detect activity if the user is beyond the camera’s field of view. Other factors such as lighting conditions, clutter, and occlusion can also adversely affect recognition performance. Consequently, these systems may not be ideal for real-time HAR or for HAR applications that require continuous operation. Over the past decade, the emergence of inexpensive, energy-efficient, and compact Inertial Measurement Units (IMUs) has been a game-changer. Market analysis indicated that the global wearable devices market will reach to about \$63 billion by 2025 \cite{zhang2022deep}. This growth in IMUs sensor technology and ubiquitous computing has not only made wearable sensor-based human action recognition (WSHAR) approaches more common but also ensured the preservation of user privacy. Currently, sensors can be embedded in a variety of portable devices, such as smartphones, smartwatches, smart cloths and other specifically-designed devices as shown in Figure \ref{figure1}.b. These devices, equipped with accelerometers and gyroscopes data, enable unobtrusive tracking of human motion and activity logging for WSHAR purposes.

Among these WSHAR systems, smartphone and smartwatch have emerged as primary platforms not only due to their embedded sensors, but also their communication, processing, and user feedback capabilities \cite{galan2019assessing}. Smartwatches, in particular,  may outperform smartphones in this regard because smartphones are only ‘on the user’ for about 23\% of the time \cite{van2015wear}, and their position relative to the user’s body is not fixed \cite{rawassizadeh2014wearables} as shown in Figure \ref{figure1}.c. Despite their advantages, WSHAR system also present certain challenges. Notably, these WSHAR systems tend to have significantly lower accuracy performance compared to systems based on visual modalities \cite{ni2022progressive}. For example, a prior study observed that fall detection, when using accelerometer data from a wrist-worn watch and processed with deep learning (DL) methods, can only attain an accuracy rate of 86\% \cite{mauldin2018smartfall}. In addition, the measurements from wearable devices are sensitive to the sensor’s location on the body. Generally, augmenting the number of sensors on various body parts (\emph{e.g.,} head, wrists, waist, legs, feet) can enhance the performance and robustness of WSHAR  systems \cite{wang2019survey}. Therefore, many existing systems based on IMUs require users to wear multiple sensors on various body parts. However, the complex deployment of multiple sensors on the body could lead to higher costs, obtrusive practical implementation challenges, and hinder the user’s ability to perform activities naturally, particularly for older users who are capable of living independently. Moreover, the combination of sensors across human body may still struggle to accurately recognize certain activities that exhibit similar sensor-derived characteristics, such as putting on jackets and falling \cite{maray2023transfer}. Consequently, effectively addressing the challenge of HAR requires an approach that goes beyond the limitations of single-modality systems.

In daily life, we engage with our surroundings through a multimodal cognitive system. This is mainly due to the fact that different regions of our brain are responsible for processing this diverse information, each specializing in a specific type of sensory input \cite{bornkessel2015neurobiological}. This system allows us to perceive objects, interpret sounds, and understand text on a daily basis. For example, reading can stimulate the reconstruction of corresponding visual imagery in our minds. Therefore, utilizing multimodal data can be advantageous in interpreting complex activities, as it encompasses a wealth of semantic knowledge \cite{sun2022human}. Meanwhile, multimodal data is a rich source of information that can reveal long-term temporal relationships between objects. These relationships are similar to the sequential order of activities within an extended sequence, much like how the human brain works \cite{dang2020sensor}. This comprehensive analysis should encompass the interpretation of objects, scenes, and the temporal relationships of activities. For instance, when a person recalls a memory, it’s like triggering a sequence in a long-term video. Such a strategy not only offers a more complete perspective but also improves our capacity to forecast activities over extended timeframes. 

\begin{figure*}[t]
  \centering
\includegraphics[width=\textwidth]{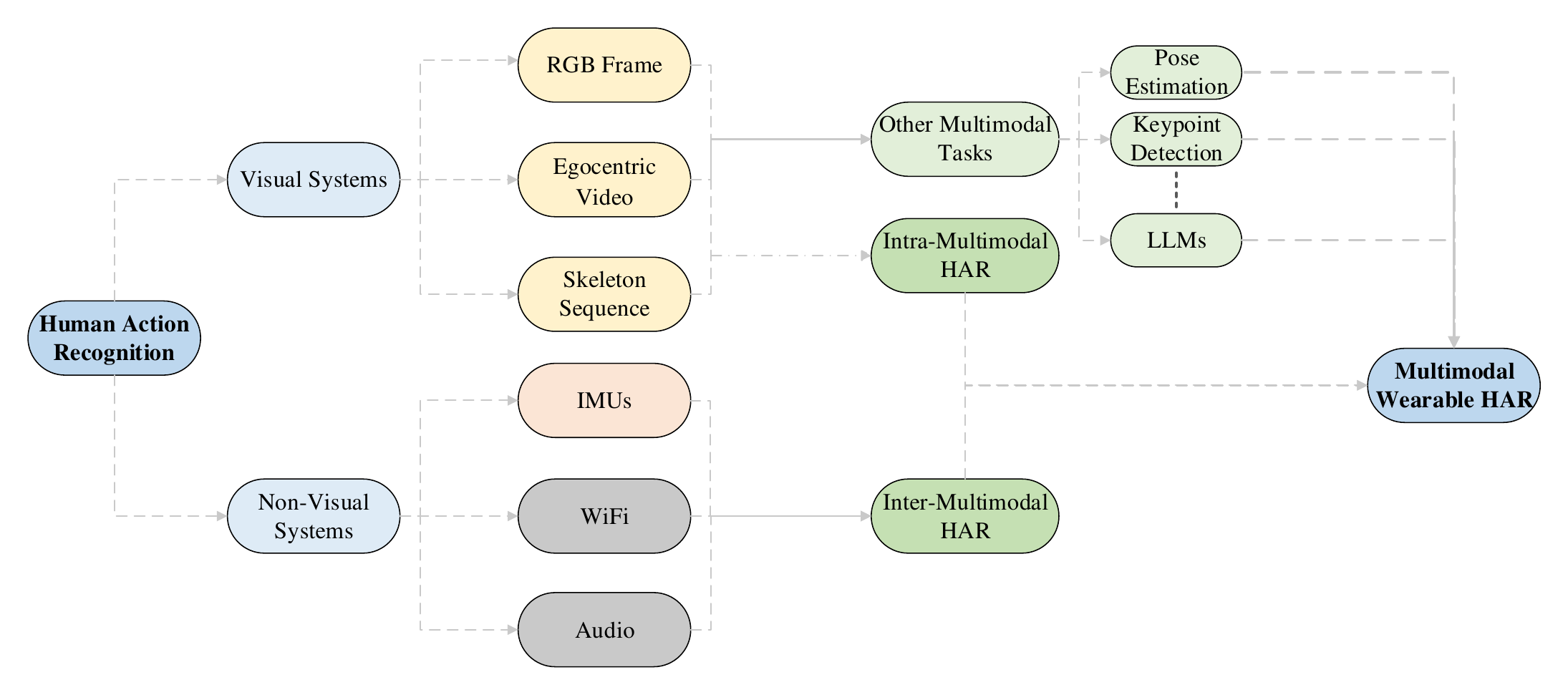}
  \caption{ Overall structure of our survey. We first present two mainstream representations available for HAR systems (Visual and Non-Visual) and their current achievements. Next, we proceed to introduce multimodal applications to emphasis on the emergence need in wearable HAR domain. We take extra efforts by combining existing multimodal studies from other tasks to form the basis for our discussions on the
existing challenges and possible future directions. }
  \label{figure2}
  \vspace{-0.3 cm}
\end{figure*}

Previously, there have been several surveys which focused on the taxonomy of HAR and reviewed HAR systems implemented with conventional machine learning methods. For instance, Poppe \cite{poppe2010survey} reviewed vision-based HAR research, discussing image representation and action classification methods. Aggarwal and Ryoo \cite{aggarwal2011human} presented a taxonomy for HAR based on their approach, discussing recognition methods for simple human actions and high-level activities. Incel \emph{et al.} provided a smartphone activity recognition taxonomy, introducing the process and challenges of HAR on phones, and reviewing works based on location, motion, and other contextual information \cite{incel2013review}. Lara and Labrador \cite{lara2012survey} surveyed HAR with wearable sensors, discussing different types of activities, design issues, and recognition methods. They evaluated 28 systems in terms of recognition accuracy, energy consumption, obtrusiveness, and flexibility. Bulling \emph{et al.} \cite{bulling2014tutorial} provided a tutorial on HAR with conventional machine learning methods based on wearable IMUs. More recently, with the development of DL methods, several state-of-the-art surveys were conducted for HAR problem \cite{wang2019survey, beddiar2020vision,yadav2021review,gu2021survey, chen2021deep, ramanujam2021human, 
sun2022human,zhang2022deep,kong2022human,qiu2022multi,dhekane2024transfer, nerella2023Transformers,yangintelligent}. However, these surveys are either focusing on advanced DL approaches \cite{dhekane2024transfer,gu2021survey,nerella2023Transformers,qiu2022multi,ramanujam2021human} or only each single sensor modality \cite{beddiar2020vision,chen2021deep,wang2019survey,zhang2022deep,yangintelligent}. While there are also a few surveys, such as \cite{yadav2021review,sun2022human}, that have summarized existing HAR methods from the perspective of data modalities, they have not emphasized how to use multimodal learning approaches to enhance WSHAR performance. As mentioned earlier, each data modality has its own strengths and limitations, and understanding how to leverage these can be crucial for improving WSHAR performance. Motivated by these observations, this comprehensive survey is designed to bridge the existing knowledge gap by connecting the advanced multimodal achievements from computer vision (CV) or natural language processing (NLP) domains with current WSHAR area.  It emphasizes the exploration of different modalities’ strengths and how they can be leveraged to enhance WSHAR’s overall performance. It serves as an invaluable resource for new researchers venturing into the field of WSHAR, providing them with a wealth of information and guidance. Those who are grappling with the choosing appropriate methods to address challenges in WSHAR field can find strong clues in this survey. The main structure is illustrated in Fig.\ref{figure2}.

The overall structure of the survey is as follows: In Section 2, we provide an overall analysis in terms of data characteristics for HAR problem, including visual representations (RGB frame, egocentric video, and skeleton sequences) and non-visual representations (audio, WiFi, and inertial sensor data). Next, we introduce current multimodal datasets which include inertial sensor data for WSHAR  problem in Section 3. We also devise the current multimodal approaches for WSHAR  from two perspectives: inter-multimodal learning (modalities from visual and non-visual systems) and intra-multimodal learning (modalities from non-visual systems only). We then proceed to present how the latest multimodal approaches can be a solution to solve some common challenges in WSHAR  domain in Section 4. In addition to the task-wise and technical introduction, we also make extra efforts by connecting the existing multimodal literature from other tasks to form the basis for possible future directions in Section 5. Section 6 includes the final remarks and conclusions. 

\section{Data Analysis for HAR}

In real life, each modality, with its distinct advantages and disadvantages, contributes to our understanding of complex systems and phenomena. In this section, we mainly focus on the inherent traits of various data modalities, including visual and non-visual based HAR systems. Specifically, visual-based modalities, such as video frame, egocentric video and skeleton sequence, which are the primary source of information in the human sensory system, is introduced first. After that, we present the current achievements using non-visual modalities for HAR problem, including audio, WiFi and inertial sensor data.

\begin{table*} [t] \small
  \caption{ Various modality examples of visual and non-visual systems with pros and cons}
  \label{tab:1}
  \centering
\begin{tabular}{c|c|c|c|c}
    \toprule
    Type & Modality  & Mainstream Methods & Pros & Cons \\ 
   \midrule  
\multirow{3}{*}{}    & & 3D CNNs  \cite{feichtenhofer2019slowfast,lin2019tsm} &  $\cdot$ Rich appearance information & $\cdot$ Viewpoint sensitive\\
                                   & RGB Frame & Transformer \cite{yang2022recurring,yan2022multiview} & $\cdot$ Easy to operate & $\cdot$ Privacy concerns \\ \cline{2-5} 
                                    Visual &   &  & $\cdot$ Active Behavior & $\cdot$ Motion alteration \\
                                       & Egocentric Video  & Transformer \cite{zhang2023helping,shiota2024egocentric} & $\cdot$ First person view  & $\cdot$  Privacy concerns \\ \cline{2-5} 
                           & & GNNs \cite{yan2018spatial,shi2019two} &  $\cdot$ Simple yet informative & $\cdot$  No appearance
                                information \\
                                    & Skeleton Sequence  & Transformer \cite{qiu2022spatio,duan2023skeletr} & $\cdot$ Viewpoint Insensitive & $\cdot$ No shape
                                information \\ \hline
\multirow{3}{*}{} &  &  & $\cdot$ Ubiquity & $\cdot$ Noise inference\\
                                    & Audio  & CNNs \cite{laput2018ubicoustics,liang2019audio}& $\cdot$  Low cost  & $\cdot$  Variability   \\
                                 \cline{2-5} 
                                Non-Visual & & & $\cdot$ Wide coverage & $\cdot$ Robustness \\
                                   &  WiFi & CNNs \cite{chen2018wifi,sheng2020deep} & $\cdot$ Penetrating ability & $\cdot$  Activity Variation \\ \cline{2-5} 
                                 &  &  & $\cdot$ Low cost & $\cdot$ less expressiveness \\
                                   & Inertial Sensor & RNNs \cite{wang2019human,xia2020lstm}& $\cdot$ Privacy preservation & $\cdot$ Noisy \\ \hline
\end{tabular}
\vspace{-0.3 cm}
\end{table*}

\subsection{Visual-based System}

\subsubsection{RGB Frame}

In the early days, handcrafted features approaches are applied to calculate the movements and spatial changes in the video to conduct HAR problem \cite{bobick2001recognition}. After that, the application of DL methods on HAR has received considerable attentions. In general, video-based HAR work can be categorized into three sections: 2D Convolutional Neural Network (CNNs), Recurrent Neural Network (RNNs) and 3D CNNs-based methods \cite{sun2022human}. For instance, Simonyan and Zisserman present a two-stream CNNs which use spatial and temporal features from individual RGB frame of video input for HAR  \cite{simonyan2014two}. RNNs was also used to analyze temporal sequence data due to the recurrent connections in their hidden layers \cite{ullah2017action}. Subsequently, several 3D CNNs approaches has been proposed recently. For instance, Feichtenhofer {\emph{et al.}} \cite{feichtenhofer2019slowfast} proposed a two-stream 3D CNNs framework consisting of a fast and slow pathway. Of this model, the slow pathway operate on the RGB frames at a low rate to capture semantic features, while the fast pathway work on high frame rates to extract motion features. In addition, Lin {\emph{et al.}} \cite{lin2019tsm} proposed a Temporal Shift Module (TSM), which shifts a part of the channels along the temporal dimension. The information from adjacent frames was then interacted with the current frame after shifting. More recently, Transformer architecture was applied to video-based HAR field \cite{yang2022recurring,yan2022multiview}. For instance, in \cite{yan2022multiview}, Multiview Transformers, which utilize multiple individual encoders, was proposed. Lateral connections between these individual encoders was integrated to efficiently fuse information from different representations of the input video. In summary, video modality contains rich RGB information and it is easy to collect for HAR problem. However, RGB video are often sensitive to various viewpoints and occlusions and are not privacy-preserving.

\subsubsection{Egocentric Video}

Egocentric videos, as opposed to third-person RGB videos, are rich in intrinsic features which are beneficial for HAR problem as these features encompass interactions with objects without occlusion \cite{li2015delving}. Currently, there are mainly two categories for egocentric HAR research area, including object-driven and motion-driven approaches \cite{nunez2022egocentric}. At present, object-driven approaches demonstrated that objects present in the scene and, specially, objects related to tasks are the main cues in the recognition of actions \cite{fathi2011understanding,aboubakr2019recognizing,nagarajan2020ego}. 
In addition to cues from objects, motion-related cues, such as the overall movement produced by scene objects, also play a crucial role for HAR problem \cite{poleg2016compact,narayan2014action}. More recently, Transformer architecture was applied to egocentric video-based HAR field \cite{zhang2023helping,shiota2024egocentric}. For instance, Zhang {\emph{et al.}} propose an object-aware decoder to enhance the performance of spatio-temporal representations for HAR using egocentric videos \cite{zhang2023helping}. This approach involves augmenting object awareness during training by training the model to predict hand positions, object positions, and semantic labels of objects using paired captions when provided. During inference, the model only needs RGB frames as inputs and can effectively track and ground objects. However, one of the primary issues using egocentric videos is the substantial camera motion resulting from the wearer's movements, which can lead to inconsistent and unstable footage. Additionally, the wearer's viewpoint, which defines the field of view in egocentric videos, may not encompass the entire context of the action, particularly if it involves other people or objects outside this field. Privacy is another concern as the camera might inadvertently capture sensitive details about the wearer or others present in the video. 

\subsubsection{Skeleton Sequence}

Skeleton sequences encode the trajectories of human body joints, which characterize informative human motions, can be another candidate for HAR problem. Previously, Various methods \cite{du2015hierarchical, zhang2017geometric} have applied RNNs and LSTMs to effectively model the temporal context information within the skeleton sequences for HAR problem. Due to the expressive power of graph neural networks (GNNs), analyzing graphs with learning models have received great attention recently \cite{yan2018spatial,shi2019two,li2021symbiotic}. For example, Yan {\emph{et al.}} exploited GNNs for skeleton-based HAR by introducing Spatial-Temporal GCNs (ST-GCNs) that can automatically learn both the spatial and temporal patterns from skeleton data \cite{yan2018spatial}. More specifically, the pose information was estimated from the input videos and then passed through the spatio-temporal graphs to achieve action representations with strong generalization capabilities for HAR. More recently, Transformer architecture was applied to HAR field using skeleton sequences \cite{qiu2022spatio,duan2023skeletr}. For instance, Qiu {\emph{et al.}} introduced the SpatioTemporal Tuples Transformer (STTFormer) architecture for HAR problem. The STTFormer initially divides a skeleton sequence into non-overlapping clips and then utilizes a spatio-temporal self-attention module to capture multi-joint dependencies between adjacent frames. Finally, an inter-frame feature aggregation module aggregates sub-actions to refine the recognition process.

In summary, the skeleton modality provides the body structure information, which is simple, efficient, and informative for representing human behaviors. Nevertheless, HAR using skeleton data still faces challenges, due to its very sparse representation, the noisy skeleton information, and the lack of shape information that can be important when handling human-object interactions. 

\subsection{Non Visual-based System}

In addition to systems based on visual data, non-visual systems have received significant attentions recently due to their robustness, ability to preserve privacy, and potential for integration into multimodal recognition systems \cite{al2020zero}. These attributes make them a valuable complement to visual systems in the field of HAR research. 

\subsubsection{Audio}

Sound serves as an effective medium for capturing the structural characteristics of human activities. Previously, this approach has proven feasible across various sensing platforms and application domains, including bathroom-related activities \cite{chen2005bathroom} and context recognition systems \cite{eronen2005audio}. Moreover, several studies applied DL methods to perform general HAR from audio signals \cite{lane2015deepear,liang2019audio,laput2018ubicoustics,yatani2012bodyscope}. For example, Lane Lane {\emph{et al.}} designed DeepEar, a pilot mobile application for multi-task sound-based detection \cite{lane2015deepear}. Yatani and Truong \cite{yatani2012bodyscope} developed the BodyScope system, capable of detecting 12 human activities related to throat movement. Laput  {\emph{et al.}} also present a a plug-and-play HAR system that leverages sound features from multiple online datasets \cite{laput2018ubicoustics}. Liang and Thomaz \cite{liang2019audio} further applied a pre-trained large-scale model to extract acoustic embedding features from public YouTube video sound clips to improve HAR performance in real-world settings. This framework integrates transfer learning, oversampling, and a deep learning architecture, eliminating the requirement for feature augmentation or semi-supervised methods. However, Using audio alone for accurate HAR is uncommon due to its limited ability to provide sufficient information. 

\subsubsection{WiFi}

WiFi is a prevalent indoor wireless signal which can be used for HAR, and even through-wall HAR, as human bodies reflect wireless signals well \cite{wu2018tw}. 
Most existing WiFi-based HAR methods focus on using Channel State Information (CSI), a fine-grained information derived from raw WiFi signals, for HAR tasks \cite{sun2022human,wang2016device}. The unique variations in CSI at the WiFi receiver are usually generated by the reflected WiFi signal of a person performing an action. 
Previously, LSTM networks have been used for HAR using CSI signal \cite{sheng2020deep,chen2018wifi}. 
In the work of \cite{sheng2020deep}, the spatial features of the CSI signal were first extracted from a fully connected layer of a pre-trained CNN. These features were then fed to a Bi-LSTM to capture the temporal information for HAR. Chen {\emph{et al.}} \cite{chen2018wifi} directly passed the raw CSI signal through an attention-based Bi-LSTM to predict the action class. Different from the above-mentioned works, Gao {\emph{et al.}} \cite{gao2017csi} transformed the CSI signal into radio images, which were then fed to a deep sparse auto-encoder to learn features for HAR problem. However, there are still some challenges which need to be further addressed, such as how to more effectively use the CSI phase and amplitude information, and to improve the robustness when handling dynamic environments.

\subsubsection{Inertial Sensor}

Rapidly development of wearable devices, such as smartwatch and smartphone, make it suitable to monitor HAR problem. Previous work has been investigated on how to apply CNN on wearable-based HAR \cite{chen2015deep}. 
RNNs type model was then suggested to deal with such time-dependent input sequences \cite{xia2020lstm, ordonez2016deep}. Wang {\emph{et al.}} \cite{wang2019human} integrated a CNN and Bi-LSTM model to acquire spatial and temporal features from acceleration data. Meanwhile, some approaches also suggested converting wearable sensor sequences as images for HAR study. 
Lu {\emph{et al.}} \cite{lu2019robust} encoded the tri-axial acceleration data into color images, which were fed into a ResNet for HAR later. Jiang and Yin \cite{jiang2015human} assembled accelerometer and gyroscope sequences into an active image. After that, a CNN model was adopted to learn the optimal features from the generated active image. Currently, there is still a significant performance gap between WSHAR and visual-based HAR systems due to intra-modality variations.

\section{multimodal Human Activity Recognition}
In real life, humans often perceive the environment in a multimodal cognitive way. By leveraging the advantages and capabilities of various data modalities, multimodal machine learning can often offer more robust and accurate HAR. In this section, we summarize the latest achievements in multimodal learning designed for current WSHAR research from two perspectives: inter-multimodal systems (utilizing modalities from both visual and non-visual systems) and intra-multimodal systems (using modalities from non-visual systems only).

\subsection{Inter Multimodal HAR Approach}

Multimodal learning is a modeling method combining and processing compensatory information from multiple modalities. Therefore, visual and non-visual modalities can be fused to leverage their complementary discriminative capabilities for more accurate and robust HAR models. Previously, some handcrafted feature-based methods \cite{chen2014improving,elmadany2018multimodal} have exploited the fusion of acceleration and visual modalities for HAR. For example, both feature and decision level fusion was used by \cite{chen2014improving}, with the combination of Kinect depth sensor and IMUs using collaborative representation classification. Their results showed 2-23\% improvement in the accuracy with the combination of depth and IMUs in comparison to the situations where they are used individually. Elmadany \emph{et al.} \cite{elmadany2018multimodal} further fused the RGB video, depth, skeleton and IMUs sensor data for addressing the HAR problem. The study employed both bimodal hybrid centroid canonical correlation analysis (BHCCCA) and multimodal hybrid centroid canonical correlation analysis (MHCCCA) to explore discriminative and informative shared spaces.

\begin{figure*}[t]
  \centering
\includegraphics[width=\textwidth]{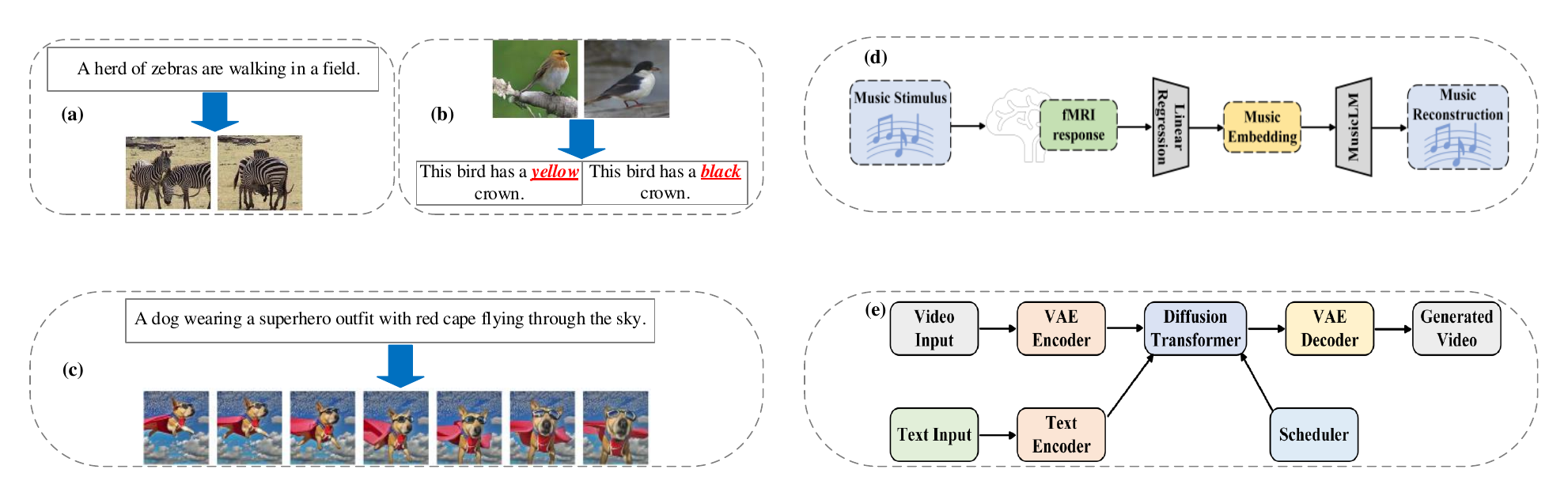}
  \caption{Current advanced multimodal tasks for other tasks. (a) Text-to-image generation task \cite{singer2022make}. (b) Image-to-text generation task \cite{venugopalan2015sequence}. (c) Text-to-image generation task \cite{qiao2019mirrorgan}. (d) Reconstructing music from human brain activity \cite{denk2023brain2music}. (e) Recent Sora study using diffusion models for video generation tasks \cite{peebles2023scalable}.}
  \label{figure3}
  \vspace{-0.3 cm}
\end{figure*}

More recently, several DL methods \cite{wei2019fusion,ahmad2019human} have been proposed for inter-multimodal HAR systems. Combining video cameras and IMUs enhances HAR performance, as IMUs typically provide orientation and acceleration information for body segments, while videos offer positional information \cite{ravi2016deep}. Marcard {\emph{et al.}} \cite{von2016human} introduced a hybrid tracker that combined multi-view video camera and IMUs for motion capture to mitigate the limitations of each sensor type. This approach used videos to obtain drift-free body position and accurate limb orientations and robust performance during rapid motions using IMUs sensors. Li \emph{et al.} \cite{li2020fusing} integrated incremental learning and decision table with swarm-based feature selection to achieve fast and accurate HAR performance by fusing kinect camera and wearable sensor data. Ranieri {\emph{et al.}} \cite{ranieri2021activity} use simultaneously data from videos, wearable IMUs and ambient sensors for HAR problem and result indicated that introduction of data from ambient sensors expressively improved the accuracy results. More recently, Ijaz \emph{et al.} \cite{ijaz2022multimodal} proposed a multimodal Transformer for nursing action recognition, where the correlations between skeleton and acceleration data are fused together. Specifically, the tokens from the temporal transformer block of the spatial-temporal skeleton model serve as queries, while the tokens from the acceleration encoder block serve as key and value pairs.

Additionally, several approaches have converted inertial signals into images to fully leverage advanced computer vision models \cite{dawar2018convolutional,wei2019fusion,das2020mmhar,qin2020imaging}. For instance, Dawar \emph{et al.} \cite{dawar2018convolutional} represented inertial signal as an image, and utilized two CNNs to fuse the depth images and inertial signals using score fusion. Wei \emph{et al.} \cite{wei2019fusion} respectively fed the 3D video frames and 2D inertial images to a 3D CNN and a 2D CNN for HAR, and the score fusion achieved better performance than feature fusion. Das {\emph{et al.}} \cite{das2020mmhar} present a multimodal ensemble networks which consisted of three models. CNNs are made for skeleton sequence while one CNN and one LSTM was trained for RGB images. After that, accelerometer and gyroscope data was converted to signal diagram for another CNN model. Qin {\emph{et al.}} \cite{qin2020imaging} also encoded time series of sensor data as images. After that, a fusion residual network is adopted by fusing these heterogeneous data with pixel-wise correspondence. In this way, WSHAR problem can be transformed as image recognition task using CV techniques.

Another type of multimodal HAR category is refereed as knowledge transfer and knowledge distillation (KD) is considered as a general technique to assist the learning process from teacher modality to student one \cite{hinton2015distilling}. Kong {\emph{et al.}} \cite{kong2019mmact} proposed a multimodal attention distillation method to model video-based HAR with the instructive side information from inertial sensor modality. Liu {\emph{et al.}} \cite{liu2021semantics} introduced the Semantics-aware Adaptive Knowledge Distillation Networks (SAKDN) model on HAR. In this model, the knowledge from multiple wearable sensors were adaptively transferred to video modality. More recently, Ni {\emph{et al.}} present the first multimodal KD approach for the WSHAR  problem \cite{ni2022cross}. In this study, an adaptive transfer of complementary information from the video domain to the sensor domain was proposed to improve the accuracy of sensor-based HAR. In order to eliminate the privacy concern, they further adopted skeleton sequence modality as the teacher model to distill knowledge to time-series modality for accurate WSHAR  problem \cite{ni2022progressive}. Of these frameworks, they will not only improve the accuracy performance of WSHAR, but also reduce the computation resource demands during the testing phase. However, a significant drawback is that student models, which take time-series data as input, typically exhibit lower accuracy performance compared to the pre-trained teacher model. This suggests that the KD method may not be able to fully exploit the advantages of multimodal learning for HAR problems due to the problem of the performance gap, which refers to the performance difference between the teacher and student models \cite{mirzadeh2020improved,tian2022learning,tian2023knowledge}. 

\begin{table*} [t] \small
  \caption{Representative multimodal benchmark datasets with various data modalities for WSHAR .  S: Skeleton, D: Depth, Au: Audio, Ac: Acceleration, Gyr: Gyroscope, EMG: Electromyography.}
  \label{tab:2}
  \centering
\resizebox{0.9\textwidth}{!}{
  \begin{tabular}{c c c c c c c c} 
    \toprule
    Dataset  & Year &  Modality & \# Class & \# Subject & \# Sample & \# Viewpoint\\
    \midrule
    Gabel \emph{et al.} \cite{gabel2012full} & 2012  & D,Ac & 6 & 23 & - & 1 \\
     Berkeley MHAD \cite{ofli2013berkeley} & 2013  & RGB,S,D,Au,Ac & 12 & 12 & 660 & 4 \\
    Delachaux \emph{et al.} \cite{delachaux2013indoor} & 2013  & D,Ac & 11 & - & - & 4 \\
    Liu \emph{et al.} \cite{liu2014fusion} & 2014  & D,Ac & 6 & 3 & - & 1 \\
    UTD-MHAD \cite{chen2015utd} & 2015  & RGB,S,D,Ac,Gyr & 27 & 8 & 861 & 1 \\
    Malleson \emph{et al.}  \cite{malleson2017real} & 2017  & RGB,Ac & - & 8 & - & 1 \\
   Dawar \emph{et al.} \cite{dawar2018action} & 2018  & D,Ac & 5 & 12 & - & 1 \\
    Manzi \emph{et al.} \cite{manzi2018enhancing} & 2018  & RGB, D,Ac & 10 & 20 & - & 1 \\
     MMAct \cite{kong2019mmact} & 2019  & RGB,S,Ac,Gyr,etc. & 37 & 20 & 36,764 & 4 \\
     EV-Action  \cite{wang2020ev} & 2020  & RGB,S,D,EMG & 20 & 70 & 7,000 & 9 \\
     HOMAGE \cite{rai2021home} & 2021  & RGB,Ac,Gyr,etc. & 75 & 27 & 1,752 & 2-5 \\
     Ego4D \cite{grauman2022ego4d} & 2022  & RGB,S,D,Au,Ac & - & 923 & - & 1 \\
    EPIC-KITCHENS-100 \cite{damen2022rescaling} & 2022 & RGB,Au,Ac  & - & 45 & 89,979 & 1 \\
    VIDIMU \cite{martinez2023multimodal} & 2023 & RGB,Ac  & 13 & 54 & - & 1 \\
  \bottomrule
\end{tabular}}
\vspace{-0.3 cm}
\end{table*}

\subsection{Intra Multimodal HAR Approach}

Currently, some studies have also combined various sensors from non-visual systems by fusing them to increase the HAR performance. For instance, CNNs was introduced to identify human activities by gathering multi-channel time-series data by employing several IMUs \cite{yang2015deep}. Chetty {\emph{et al.}} \cite{chetty2015intelligent} presented a multimodal CNNs system using gyroscope and accelerometer data from smartphone, applying it to eHealth scenarios for the elderly and people with special needs. 
In \cite{guo2016wearable}, the authors integrated various IMUs using a feature ensemble method from multiple wearable sensors. Additionally, they introduced a layered fusion model that utilizes entropy weight to track human activities using these IMUs \cite{guo2018multisensor}. Yao {\emph{et al.}} designed a architecture which consist of three different CNNs sequential blocks which can learn local patterns, high-level relationship as well as temporal features among input sensors, to merge multimodal data for sensor-based HAR problem \cite{yao2017deepsense}. After that, Sena {\emph{et al.}} utilized multi-scale CNNs ensemble approach to not only extract both simple movement patterns as well as complex movements to deal with data heterogeneity problem \cite{sena2021human}.

LSTM-based model was designed in \cite{ullah2019stacked}, where the data obtained from the gyroscope and accelerometer were first normalized. After that, the normalized data were then passed on to the stacked LSTM network for HAR problem. Yu {\emph{et al.}} \cite{yu2018human} proposed a Bi-LSTM model utilizing data from gyroscope and accelerometer sensors from a mobile phone for HAR problem. Moreover, Ihianle {\emph{et al.}} \cite{ihianle2020deep} present a multi-channel architecture using both CNNs and BLSTM to extract features from multimodal sensing devices for activity recognition. Dua {\emph{et al.}} \cite{dua2021multi} further integrated CNNs with Gated Recurrent Unit (GRU) modules to extract long-term temporal dependencies from accelerometer and gyroscope sensors data for HAR problem.  

More recently, attention mechanism was introduced for multimodal sensing HAR \cite{gao2021danhar,tang2022triple,al2022multi}. For instance, Gao {\emph{et al.}} present a dual attention framework that integrates both channel and temporal attention simultaneously to capture temporal-spacial patterns from accelerometer and gyroscope for HAR task  \cite{gao2021danhar}. Tang {\emph{et al.}} \cite{tang2022triple} further proposed a a triplet cross-dimension attention for WSHAR problem. These three attention branches were used to capture the cross-interaction between sensor dimension, temporal dimension and channel dimension from accelerometer and gyroscope sensor signals. Al-qaness {\emph{et al.}} incorporated RNNs with attention module to extract time-series feature for wearable HAR problem \cite{al2022multi}. 

Simultaneously, other works explore non-visual modalities, such as audio or WiFi signals, in conjunction with accelerometer data for multimodal HAR problems \cite{zhang2017fingersound,garcia2018multi,siddiqui2020multimodal,qiu2022multi}. 
For instance, Garcia {\emph{et al.}} \cite{garcia2018multi} treated audio and accelerometer sensor data as different views and applied stacked generalization approach to fuse them for wearable HAR problem. Siddiqui and Chan \cite{siddiqui2020multimodal} further investigated the use of acoustic signals with an accelerometer and gyroscope from the human wrist for gesture recognition. Mollyn {\emph{et al.}} first employed the IMU inputs to serve as a trigger for identifying activity events. Upon detection of these events, a multimodal learning model which augmented the IMU samples with sub-sampled audio data capture from a smartwatch was proposed for WSHAR problem \cite{mollyn2022samosa}. In addition, Lin {\emph{et al.}} \cite{lin2022human} proposed a multimodal system using a smartphone with an off-the-shelf WiFi router for HAR problem. The router functions as a hotspot for transmitting WiFi packets, while the smartphone is equipped with customized firmware and developed software to capture WiFi CSI information simultaneously.

While there are numerous studies focusing on multimodal learning approaches for WSHAR problems, the current methods employed in WSHAR field are still in their infancy compared to the models used in other advanced tasks, such as text to image/video generation \cite{singer2022make,qiao2019mirrorgan,peebles2023scalable} or music reconstruction from brain activity \cite{denk2023brain2music} as illustrated in Figure \ref{figure3}. More recently, Large Language Models (LLMs) have gained substantial attentions since preliminary results indicated that LLMs possessed distinct capabilities in utilizing inherent world knowledge to interpret IoT sensor data and make logical deductions about physical world tasks \cite{xu2023penetrative}. This not only paves the way for new applications of LLMs beyond conventional text-based tasks, but also introduces innovative methods for integrating human knowledge into real-world systems. As a consequence, we highlight significant multimodal approaches addressing challenges in current WSHAR systems. Some of these solutions involve the application of multimodal learning, derived from other advanced tasks, to enhance the performance of WSHAR systems.

\begin{figure*}[t]
  \centering
\includegraphics[width=\textwidth]{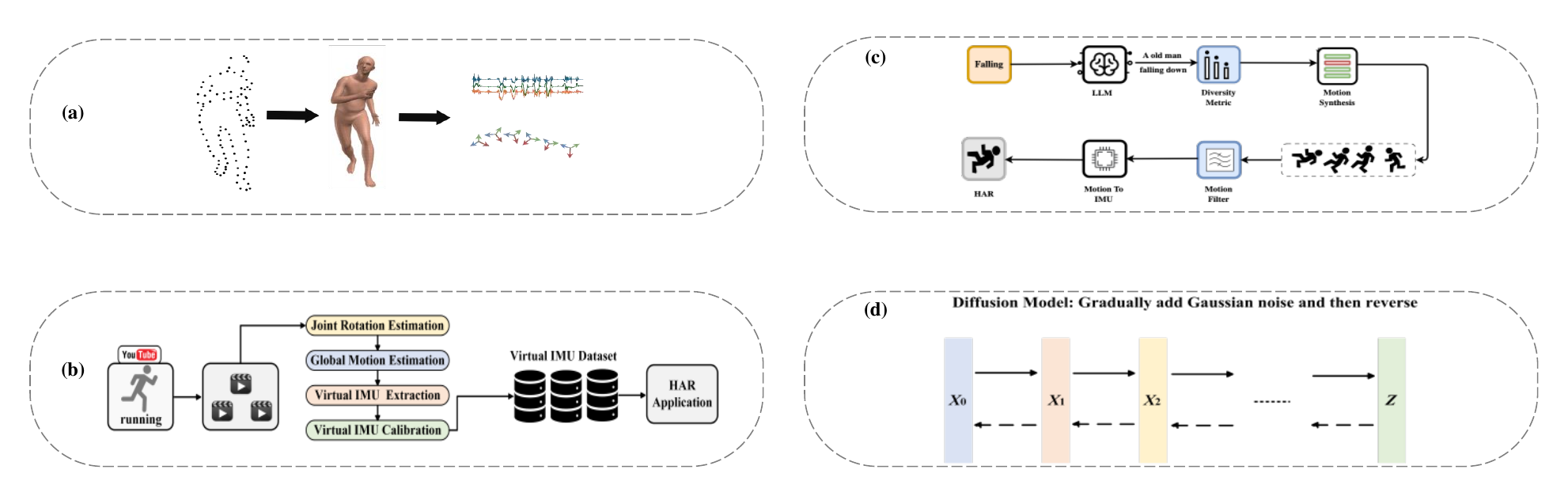} 
  \caption{Current approaches for multimodal WSHAR dataset scarcity problem. (a) Synthesized IMU data from motion capture datasets using skinned multi-person linear (SMPL) model \cite{huang2018deep}. (b) Virtual IMU data generation pipeline from LLMs domain \cite{leng2024imugpt}. (c) Virtual IMU data generation pipeline from video domain \cite{kwon2020imutube}. (d) IMU data generation using advanced diffusion model \cite{peebles2023scalable}.}
  \label{figure4}
  \vspace{-0.3 cm}
\end{figure*}

\section{Challenges in multimodal Wearable HAR System}

Previously, many studies have pointed out current challenges in the WSHAR area, including annotation scarcity, class imbalance, distribution discrepancy and computational cost \cite{chen2021deep,sun2022human}. However, the application of multimodal learning approaches as to address these challenges in WSHAR problems has not been previously explored. In the following section, we will outline some of the existing challenges in the WSHAR domain and provide recent multimodal solutions for addressing these common challenges. 

\subsection{Multimodal Dataset Challenge}

The dataset plays a critical key in the success of deep learning research. Consequently, numerous multimodal datasets have been collected to train and evaluate the HAR performances \cite{sun2022human}. However, only a few of them include IMUs data, which is essential for WSHAR problem. Additionally, there are also some other challenges, such as limited labeled data and class imbalance, existing in these multimodal datasets.

\subsubsection{Large-scale Dataset Scarcity}

In the age of deep neural networks, the absence of large-scale labeled datasets makes it challenging to fully utilize the advantages of DL methods. As a result, the development of efficient methods for acquiring labeled datasets has been a long-standing research interest across various communities. In fact, prior research has suggested that the lack of extensive dataset is a contributing factor to the slower progress of WSHAR area compared to other fields \cite{chen2021sensecollect}. Nevertheless, there are only a fewer multimodal datasets including IMUs signals for WSHAR problems, as shown in Table \ref{tab:2}. In general, MMAct \cite{kong2019mmact}, and EPIC-KITCHENS-100 \cite{damen2022rescaling} are large benchmark datasets suitable for multimodal learning approaches for WSHAR field. In addition, Berkeley MHAD and UTD-MHAD \cite{chen2015utd} datasets are also popular used. Compared to benchmark datasets from other modalities domain, such as ImageNet \cite{deng2009imagenet}, VidProM \cite{wang2024vidprom}, Panda-70M \cite{chen2024panda}, and BEHAVIOR-1K \cite{li2024behavior} in CV area, and Aya \cite{singh2024aya} in NLP domain, which contained millions of samples, current large-scale multimodal datasets for WSHAR problem are extremely under-explored. Moreover, the collection of even unlabeled sensor data presents its own set of logistical and ethical hurdles. These include the installation of sensors, the recruitment of human participants, and the need to address privacy concerns. Consequently, the development of efficient and effective methods for rapidly acquiring large-scale, fully labeled datasets would be a valuable contribution to the WSHAR research community. 

Existing studies have utilized various tasks, such as human pose estimation \cite{huang2018deep,xia2021learning,hao2022cromosim} and image classification \cite{ataseven2022physical,hashim2022deep} to leverage along with the WSHAR research. For instance, Huang {\emph{et al.}} \cite{huang2018deep} synthesized IMU data from motion capture datasets using skinned multi-person linear (SMPL) model surface \cite{loper2023smpl} as shown in Figure \ref{figure4}.a. Specifically, to generate synthetic IMU training data, virtual sensors are positioned on the SMPL mesh surface. Orientation readings are obtained directly through forward kinematics, while accelerations are derived using finite differences. Xia {\emph{et al.}} also created a comprehensive synthetic HAR dataset using the SMPL model. This dataset includes multimodal data, such as acceleration and angular velocity, which were generated according to the forward kinematics approach \cite{xia2021learning}. Hao {\emph{et al.}} also presented a high-precision virtual IMU sensor simulator from either motion capture systems or single-lens RGB cameras using the SMPL model surface \cite{loper2023smpl}. In order to reduce measurement noise and calibration errors, the functional mapping from imperfect trajectory estimations was learned by a DNN model to mitigate the data scarcity problem. Additionally, Hashim and Amutha \cite{hashim2022deep} transformed accelerometer and gyroscope sensor data to the visual image using novel activity image creation (NAICM) method. After that, pre-trained models on ImageNet \cite{deng2009imagenet} are transferred for HAR problem. Yoon \cite{yoon2022img2imu} also converted IMU sensor data into visually interpretable spectrograms. Pre-trained representations from the ImageNet \cite{deng2009imagenet} dataset were employed for diverse few-shot IMU tasks by using contrastive learning.

In situations where video and IMU data streams cannot be accessed simultaneously, a straightforward solution could be to leverage large-scale datasets from existing multimodal HAR datasets. The objective here is to utilize the data from existing HAR datasets, which are extensive and modality-agnostic, to address the issue of data scarcity in the WSHAR domain. For instance, numerous studies in the CV domain have focused on advanced methods to extract skeleton joints data from video streams \cite{morais2019learning,duan2022revisiting}. Since accelerometer data can be regarded as the second derivative of the skeleton sequence coordinates \cite{ni2023physical}, extracting accelerometer data directly from video streams is practical. Indeed, the use of such sophisticated tools has significantly advanced the field of WSHAR by generating virtual IMU data from the video domain \cite{kwon2020imutube, kwon2021approaching,yoon2022img2imu,jain2022effectiveness,gavier2023virtualimu, li2023signring,santhalingam2023synthetic}. For instance, Kwon {\emph{et al.}} proposed an engineering pipeline, IMUTube, to generate on-body virtual sensor data using data from video modality \cite{kwon2020imutube} as shown in Figure \ref{figure4}.b. The proposed processing pipeline transforms the video data into usable virtual sensor (IMU) data. This involves extacting 2D pose information from videos, which is subsequently converted to 3D. By tracking individual joints of the resulting 3D poses, sensor data like tri-axial acceleration values is generated across various locations on the body. These values are then post-processed to align with the target application domain. They tested their approach in a realistic gym exercises scenario involving large body movements. The result demonstrated that HAR systems trained with virtual sensor data significantly outperform baseline models trained only with real IMU data \cite{kwon2021approaching}.
Jain {\emph{et al.}} \cite{jain2022effectiveness} further evaluated the effectiveness of the IMUTube pipeline in detecting subtle motion activities, particularly focusing on eating detection tasks. They found that IMUTube significantly improved recognition accuracy in these tasks. Leng {\emph{et al.}} designed a motion subtlety index to measure local pixel movements and changes in pose at specified virtual sensor locations, exploring its relationship with the accuracy of activity recognition to evaluate the IMUTube pipeline \cite{leng2023generating}. Additionally, Fortes {\emph{et al.}} first trained a general regression model for both accelerometer and gyro signals \cite{fortes2021translating}. This model is then applied to video footage of specific activities, enabling the generation of synthetic IMU data that can be used to improve HAR models. In order to further improve the quality of virtual IMU data, some studies aim to produce more realistic and diverse IMU signals \cite{xia2022virtual,gavier2023virtualimu}. For instance, Gavier {\emph{et al.}} \cite{gavier2023virtualimu} propose a systematic approach to synthesize realistic and diverse IMU data, including three-axis accelerometer and gyroscope measurements, from video-based skeleton representations. 

Meanwhile, with the recent advancements of large language models (LLMs), pre-trained LLMs models that can be adapted to solve a wide range of multimodal downstream tasks  \cite{zhou2023comprehensive}. For instance, CLIP was developed to learn the association between images and their textual descriptions \cite{radford2019language}. More recent developments, such as Next-GPT \cite{wu2023next}, have pushed these boundaries even further, enabling the integration of multiple diverse modalities. Inspired by these achievements, scholars are currently investigating the capabilities of LLMs in the field of time series analysis. In order to address the issue of scarce annotated data, Li {\emph{et al.}} employed clinical reports that are automatically generated by LLMs to serve as a guide for a self-supervised pre-training framework for ECG data \cite{li2024frozen}. A trainable ECG encoder and a fixed language model were employed to embed paired ECG and automatically generated clinical reports independently. Similarly, Liu {\emph{et al.}} \cite{liu2023large} provided a systematic demonstration of how LLMs can effectively interpret numerical time series data using few-shot prompt tuning. Zhang {\emph{et al.}} evaluated the proficiency of LLMs, such as Claude-2 \cite{bai2022training}, in identifying unusual patterns in mobility data \cite{zhang2023large}. The experimental observations indicated that LLMs are capable of achieving commendable performance in anomaly detection. Liu {\emph{et al.}} further demonstrated that LLMs are capable of grounding time series data for activity recognition with only few-shot tuning approaches \cite{liu2023large}. 

More recently, a few studies started to use LLMs to generate varied virtual IMU data for a wide range of real-world activity contexts \cite{leng2023generating,leng2024imugpt}. For instance, Leng {\emph{et al.}} utilized LLMs to generate prompts which can subsequently processed by CLIP-based model to produce 3D human motion sequences which were converted into streams of virtual IMU data for further HAR problem \cite{leng2023generating} as shown in Figure \ref{figure4}.c. After that, virtual IMU data can be calculated using inverse kinematics based on these motion sequences. Leng {\emph{et al.}} \cite{leng2024imugpt} further proposed language-based cross modality transfer system for HAR problem. Specifically, an LLMs model generate textual descriptions of activities automatically, which which are then converted into motion sequences by a
motion synthesis model. After that, a motion filter was designed to screen out incorrect sequences to obtain only relevant motion sequences for virtual IMU data extraction. A diversity metric was introduced to measure the distribution shift between textual descriptions and motion sequences to determine when data generation should be
stopped for most effective HAR performance.

In addition to the aforementioned approaches, another solution is to use adversarial network to synthesize data. Currently, there are many studies using VAE \cite{kingma2013auto}, GAN \cite{goodfellow2014generative} and diffusion \cite{ho2020denoising,aggarwal2011human,aggarwal2021dance2music} methods for multimodal data generation purpose. For instance, Aggrawal {\emph{et al.}} \cite{aggarwal2021dance2music} present both offline and online approaches to generate music from video. Currently, there are several studies trying to use adversarial networks to produce synthetic time-series data for HAR problems \cite{wang2018sensorygans,siyal2020human,li2020activitygan,wang2022wearable,li2022tts,peebles2023scalable} as shown in Figure \ref{figure4}.d. For instance, GANs can augment smaller datasets by generating new, previously unseen data \cite{li2022tts}. More recently, Ni {\emph{et al.}} \cite{ni2023physical} proposed a cross-modal adversarial framework to produce corresponding synthetic skeleton joints from accelerometer data. Thus, wearable devices are capable of not only collecting time series data but also able to generate skeleton sequences for further multimodal process. This addresses the need for real-time applications of multimodal wearable HAR system that can be conducted ubiquitously. While adversarial networks have shown success in generating single time series, their potential in multimodal HAR system on wearable sensors remains largely untapped.

\begin{figure*}[t]
  \centering
\includegraphics[width=\textwidth]{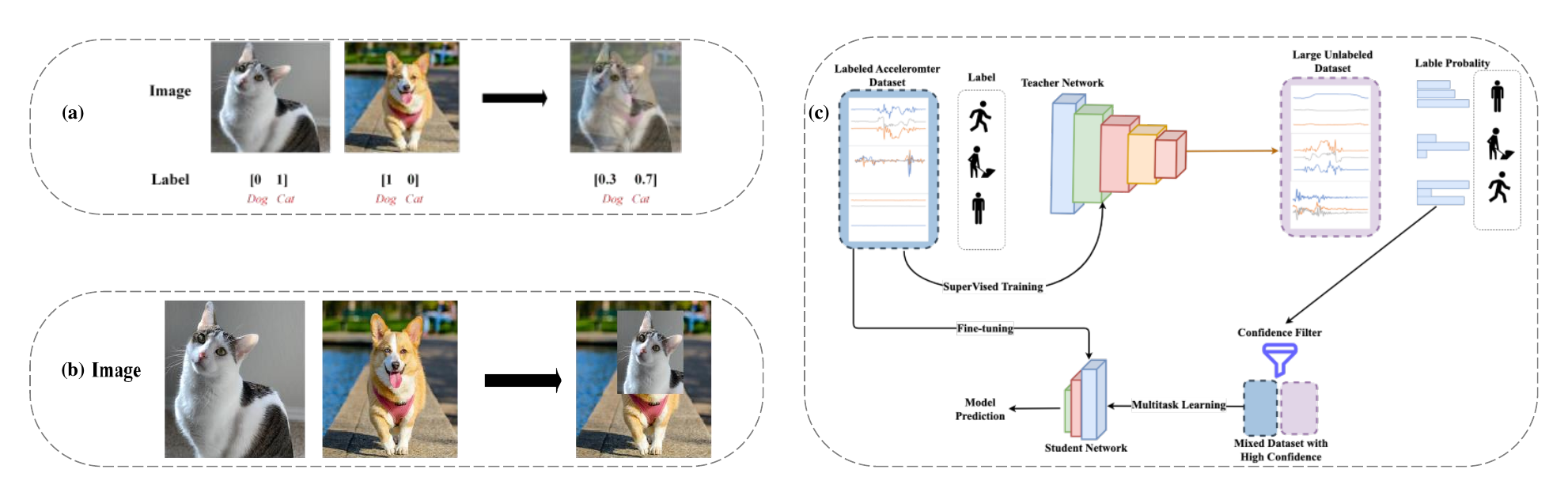}
  \caption{Current approaches for limited labeled data problem. (a) MixUp method \cite{zhang2017mixup}. (b) CutMix approach \cite{yun2019cutmix}. (c) Self-training based SSL
method \cite{tang2021selfhar}.}
  \label{figure5}
  \vspace{-0.3 cm}
\end{figure*}

\subsubsection{Limited Labeled Data in Existing Dataset}

Unlike images or audio, where annotations can be derived directly from the raw data, annotating sensor data is a complex task for humans without the aid of post-experimental video recordings. In addition, current methods for obtaining labeled datasets often require significant human effort \cite{zhang2020deep}. Therefore, current multimodal datasets for wearable HAR problem are considered as small- or medium- scale datasets. As a result, how to fully utilized such small/medium datasets is also critical for wearable HAR community. 

A straightforward solution could be leveraging data augmentation approaches from existing WSHAR tasks. Data augmentation (DA) is a method used to enhance the variety of training samples without the need to gather new data. Currently, transformations in the time series domain are among the simplest and most effective data augmentation techniques for time series data. The majority of these methods directly alter the original input time series, such as time slicing window, jittering, scaling, rotation, permutation, channel permutation and so on \cite{iglesias2022data}. At the same time, there are many data augmentation approaches including AutoAugment \cite{cubuk2018autoaugment}, MixUp \cite{zhang2017mixup} and CutMix \cite{yun2019cutmix} in CV domain as shown in Figure \ref{figure5}.a and \ref{figure5}.b. Moreover, there are many studies in multimodal learning are using data augmentation methods for various tasks \cite{renduchintala2018multi,falcon2022feature,oneata2022improving,xu2020mda,liu2022learning,hua2023multimodal,josi2023multimodal}. For instance, Renduchintala {\emph{et al.}} proposed a multimodal data augmentation network for automatic speech recognition. This network, comprising two distinct encoders, is capable of handling multimodal inputs, both acoustic and symbolic. It facilitates the effortless integration of extensive text datasets with considerably smaller transcribed speech corpora during the training process. This approach optimizes the use of available resources, enhancing the learning process. \cite{renduchintala2018multi}. Xu {\emph{et al.}} introduced a multimodal data augmentation framework aimed at enhancing performance in multimodal image-text classification tasks. The framework was designed to learn a cross-modality matching network, which selects image-text pairs from existing unimodal datasets to create a synthetic multimodal dataset. This dataset is then used to improve the performance of classifiers \cite{xu2020mda}. Falcon {\emph{et al.}} proposed a multimodal data augmentation approach using feature space and generate new videos and captions by blending samples that are semantically alike to solve the  text-video retrieval problem \cite{falcon2022feature}. Liu {\emph{et al.}} further present a method for learning multimodal data augmentation. This method can autonomously learn to augment multimodal data in the feature space, without any restrictions on the types of modalities or their interrelationships \cite{liu2022learning}. Oneata {\emph{et al.}} employed data augmentation techniques to prompt the multimodal system to focus on visual cues. The experiment results indicated that this technique is not only conceptually more straightforward but also consistently enhances performance in a multimodal environment \cite{oneata2022improving}. Hua {\emph{et al.}} designed a BERT-based back-Translation Text and Entire-image multimodal model to detect fake news. The proposed framework applied data augmentation method to not only mitigate the issue of limited data, but also generates positive samples that are beneficial for the following contrastive learning module \cite{hua2023multimodal}. Josi {\emph{et al.}} present a data augmentation method which relied on local occlusions and global modality masking methods for person re-identification problem \cite{josi2023multimodal}. While data augmentation have shown success in improving single modality HAR and other multimodal problems, its potential in leveraging multimodal approaches on WSHAR system remains significantly unexplored.

Another approach is to use self-supervised learning (SSL) to address the problem of limited labeled data. SSL is an effective technique where the model can learn good data representations from unlabeled data to improve downstream task performances. For instance, video data typically encompasses several modalities such as visuals, sound, and text. The temporally synchronized characteristics of these modalities offer an inherent method for deriving positive pairs that share the same time frame, eliminating the necessity for preliminary text tasks \cite{wang2021fine}. Consequently, many SSL methods have been initially developed for audio-visual learning \cite{asano2020labelling}, speech-visual learning \cite{miech2020end} and other tasks \cite{alayrac2020self}. For instance, Asano {\emph{et al.}} introduced a clustering method that enables pseudo-labeling of a video dataset without the need for human annotations. This is achieved by capitalizing on the inherent relationship between the audio and visual modalities \cite{asano2020labelling}. Alayrac {\emph{et al.}} \cite{alayrac2020self} proposed a multimodal versatile network which applied SSL to learn representations by leveraging visual, audio and language streams modalities from video. Specifically, they investigated the optimal way to merge these modalities, ensuring that detailed representations of both visual and audio modalities are preserved, while simultaneously incorporating text into a unified embedding. Similarly, Li {\emph{et al.}} presents a distinctive cross-modal self-supervised learning technique that models the certainty of audio and visual observations by taking advantage of the complementarity and uniformity among various modalities \cite{li2022multi}. 

Currently, there are also a few studies using SSL for either single modality WSHAR \cite{saeed2019multi,tang2021selfhar,yuan2022self,jain2022collossl} or multimodal \cite{akbari2021vatt,choi2023multimodal,brinzea2022contrastive,deldari2023latent} HAR problem. For instance, Saeed {\emph{et al.}} utilized a SSL method to learn feature from unlabeled data for HAR task. Specifically, a multi-task temporal convolutional network was trained to differentiate various transformations which performed on the raw input signals \cite{saeed2019multi}. Tang  {\emph{et al.}} proposed a self-training based SSL method for WSHAR as shown in Figure \ref{figure5}.c. Initially, a teacher model was employed to distill the knowledge from labeled accelerometer data to label a large unlabeled dataset. High-confidence data points from this step were selected, and a student model was trained to discriminate these selected activities. The ground truth labels from the training labeled dataset were further used to fine-tune the student model \cite{tang2021selfhar}. Yuan {\emph{et al.}} further leveraged SSL for one of the largest unlabeled wearable sensor dataset, UK-Biobank activity tracker dataset, for WSHAR problem \cite{yuan2022self}. Jain {\emph{et al.}} presented a collaborative SSL technique which takes advantage of unlabeled data gathered from various wearable devices used by an individual to learn superior characteristics of the data \cite{jain2022collossl}. In addition, Akbari {\emph{et al.}} presents a Transformer-based framework which takes raw signals as input and extracts multimodal representations that are rich enough to benefit a variety of downstream tasks. This framework was trained using multimodal contrastive losses and the result demonstrated its robust performance by action recognition task \cite{akbari2021vatt}. After that, Choi {\emph{et al.}} further proposes a hard negative sampling method for multimodal HAR which relies on a hard negative sampling loss for skeleton and IMUs data pairs \cite{choi2023multimodal}. Brinzea {\emph{et al.}} implements a multimodal SSL framework which can exploit modality-specific knowledge to encode inertial and skeleton data for HAR problem \cite{brinzea2022contrastive}. Deldari {\emph{et al.}} introduces a cross-modal SSL which can generate masking intermediate embeddings by modality-specific encoders. After that, these embeddings can integrated into a global embedding via a cross-model aggregator for HAR task \cite{deldari2023latent}. 

More recently, a few studies tried to apply CLIP-based framework for HAR problem \cite{radford2019language,moon2022imu2clip,xia2024ts2act}. Girdhar {\emph{et al.}} present a framework to learn a joint embedding across various modalities including images, text, audio, depth, thermal and IMUs data. This framework which uses CLIP-like architecture, can extend to new modalities just by using their natural pairing with images \cite{girdhar2023imagebind}. Moon {\emph{et al.}} proposed a contrastive learning approach that transforms the IMUs sensor readings and the textual annotations upon the videos of human activities into a shared embedding space to enhance information retrieval \cite{moon2022imu2clip}. More recently, Xia {\emph{et al.}} designed a cross-modal co-learning method for few-shot HAR problem. This method first utilized the semantic-rich label text to search for human activity images to form an augmented dataset consisting of partially-labeled time series and fully-labeled images. A pre-trained CLIP image encoder was used to train a time series encoder with constrative learning. After that, the feature extracted from the input time series is compared with the embedding of the pre-trained CLIP text encoder using prompt learning and the best match is output as the HAR classification results \cite{xia2024ts2act}. The experimental results demonstrated that the proposed method performed close to or better than the fully supervised methods even using limited labeled samples.

\subsubsection{Imbalance Class Issue}

The balanced distribution of classes in the training dataset is a fundamental assumption in the creation of many machine learning algorithms. Yet, this is not always the case, and an imbalance in class distribution could potentially introduce a bias that adversely affects the performance of these models. For example, it is difficult to collect datasets of actual unexpected falls due to the infrequency and varied circumstances of falls in real life. The problem is further amplified in multimodal datasets, where there could be numerous interrelations between majority and minority classes.

Generally, The issue of class imbalance in data can be tackled by data and algorithm levels. The objective of at the data level is to adjust the original datasets through resampling in order to achieve a balanced class distribution. This approach involves several different forms of resampling methods, such as undersampling the majority class or oversampling the minority class. For example, random oversampling is able to duplicate random instances until a certain class balance is reached \cite{sleeman2022multimodal}. However, this method might carry the potential risk of eliminating valuable instances or resulting in a final training dataset that is considered too small. At the same time, synthetic minority over-sampling technique (SMOTE) method is another popular framework at the data level \cite{chawla2002smote}. SMOTE aims to generate new instances by combining nearby instances of the same class and has demonstrated to surpass random sampling techniques in many instances \cite{sleeman2022multimodal,junaid2023explainable}. For instance, Junaid {\emph{et al.}} compared SMOTE and its variants, such as SVMSMOTE, BorderlineSMOTE, ADASYN (Adaptive Synthetic), SMOTENC (Synthetic Minority Over-sampling Technique for Nominal and Continuous), and SMOTEENN (Synthetic Minority Over-sampling Technique with Ensemble of Neighbors), to evaluate the effectiveness solving imbalance class problem. After conducting multiple experiments on various data modalities, they concluded that SMOTENN was the best fit for addressing the class imbalance in the training set for early detection of Parkinson's disease \cite{junaid2023explainable}. 

More recently, GAN approaches have been used as an oversampling methods \cite{douzas2018effective,lee2022boundary}. For instance, Lee {\emph{et al.}} presented a boundary-focused GAN (BFGAN) oversampling technique aimed at selectively controlling the placement of generated samples to tackle the class imbalance problem in multimodal time series classification \cite{lee2022boundary}. The proposed BFGAN incorporated a specifically designed additional label to reflect the importance of a sample's position in the data space. After considering both the multimodality and importance of a sample, the BFGAN generated synthetic samples using GAN approaches. Similarly, Li {\emph{et al.}} utilized a GAN approach for synthesizing samples, even when all multimodal features are missing, to address the issue of imbalanced multimodal data \cite{li2020deep}. Currently, there are a few studies trying to address class imbalance issue in the HAR field \cite{guo2021evolutionary,garcia2021distillation}. Guo {\emph{et al.}} introduce a dual-ensemble class imbalance learning method. In this method, an internal ensemble learning model which include several heterogeneous sub-classifier was designed. The one with the  highest recognition accuracy is selected as the base classifier. Sequentially, multimodal evolutionary algorithms were presented to find the optimal combination that contains the smallest number of base classifiers while accurately identifying human actions \cite{guo2021evolutionary}. Furthermore, the proposed distillation multiple choice learning framework by Garcia {\emph{et al.}} 
addresses the HAR problem by enabling different modality networks to learn cooperatively from scratch. This cooperative learning approach leads to significantly higher accuracy compared to training the networks separately, as each modality benefits from the complementary information offered by the multimodal data \cite{garcia2021distillation}.

\subsection{Heterogeneous Feature Alignment Challenge}

Typically, single modality representation involves a linear or nonlinear mapping of an individual input stream (\emph{e.g.,} image, video, or sound, etc.) into a high-level semantic representation. Multimodal representation, however, combines the correlation power of each single modality sensation by aggregating their spatial outputs. Despite this, current DL models often struggle to accurately represent the structure and representation space of both the source and target modality. In this section, we will delve into insight solutions aimed at addressing the challenges of feature alignment in multimodal HAR systems, providing a comprehensive understanding of this complex domain.

\subsubsection{Cross-model Transfer Learning}

One of the solutions for feature alignment is transfer learning. However, transfer learning heavily relies on whether the underlying domains or tasks across the source and target domains are the same \cite{dhekane2023much}. Consequently, many studies worked on heterogeneous transfer learning which refers to the case where the source and target feature spaces differ for HAR problem \cite{wei2016instilling,wang2018deep,wang2018stratified,lu2021cross,yuan2021multimodal,sung2022vl,khaertdinov2022temporal,geng2022multimodal,lu2022semantic,thukral2023cross,bao2023survey}. For instance, Wei {\emph{et al.}} proposed a co-regularized heterogeneous transfer learning model, which built a common semantic space derived from social media and labeled physical sensor data \cite{wei2016instilling}. Wang {\emph{et al.}} present an unsupervised source selection algorithm for HAR problem. The most similar k source domains from a list of available domains was selected first. After that, the time and spatial relationship between activities were captured using a transfer neural network to perform knowledge transfer for activity recognition \cite{wang2018deep}. Wang {\emph{et al.}} designed a stratified transfer learning method which can dramatically improve the classification accuracy for cross-domain activity recognition. Specifically, it first utilizes majority voting technique to capture pseudo labels from the target domain. After that, it performed intra-class knowledge transfer iteratively to transform both domains into the same subspaces. The labels of target domain are eventually obtained via the second annotation for final transfer learning stage \cite{wang2018stratified}. Qin {\emph{et al.}} present an adaptive spatial-temporal transfer learning method to adaptively evaluate the relative importance between the marginal and conditional probability distributions in spatial features. It also adopted an incremental manifold learning to capture temporal features for cross-dataset activity recognition \cite{qin2019cross}. Lu {\emph{et al.}} introduced an optimal transport-based method to better utilize the locality information of activity data for cross-domain HAR for accurate and efficient knowledge transfer \cite{lu2021cross}. It utilized clustering methods to capture the substructures of activities and sought the coupling of the weighted substructures between different domains. 

More recently, Yuan {\emph{et al.}} designed a multimodal contrastive training method for visual representation learning. It exploited intrinsic data properties within each modality and semantic information from cross-modal correlation simultaneously, hence improving the quality of learned visual representations \cite{yuan2021multimodal}. Sung {\emph{et al.}} built an adapter-based parameter-efficient transfer learning techniques for vision-and-language tasks. It employed a unified format and architecture to solve the tasks in a multi-tasking learning setup \cite{sung2022vl}. Khaertdinov {\emph{et al.}} applied a dynamic time warping (DTW) algorithm in a latent space to force features to be aligned in a temporal dimension \cite{khaertdinov2022temporal}. Geng {\emph{et al.}} proposed a simple and scalable multimodal masked autoencoder architecture to learn a unified encoder for both vision and language data via masked token prediction. The experiment results indicated that this architecture is able to learn generalizable representations that transfer well
to downstream classification tasks \cite{geng2022multimodal}. Lu {\emph{et al.}} designed a semantic-discriminative Mixup approach which considers the activity semantic ranges to overcome the semantic inconsistency brought by domain differences for generalizable cross-domain HAR problem. In addition, they also introduced the large margin loss to enhance the discrimination of Mixup to prevent misclassification brought by noisy virtual labels \cite{lu2022semantic}. Thukral {\emph{et al.}} introduces a cross-domain HAR transfer learning framework which follows the teacher-student self-training paradigm to more effectively recognize activities with very limited label information.  It can bridge conceptual gaps between source and target domains, including sensor locations and type of activities \cite{thukral2023cross}.

\subsubsection{Multimodal Knowledge Distillation}

Another solution to solve the heterogeneous feature alignment issue is to use KD method, which tries to transfer knowledge from a complicated pre-trained network ({\emph{i.e.}, teacher model) to a smaller network ({\emph{i.e.}, student model) by minimizing the Kullback-Leibler (KL) divergence of predictions between teacher and student models \cite{hinton2015distilling}. By mimicking the accuracy performance from teacher model, the student model can eventually improve its performance \cite{hinton2015distilling}. At present, there are several studies applying KD for multimodal learning tasks\cite{hoffman2016learning,garcia2018modality,andonian2022robust}. For instance, Hoffman {\emph{et al.}} proposed a modality hallucination architecture that uses depth as side information to guide an RGB object detection model \cite{hoffman2016learning}. In order to facilitate feature alignment, same multi-layer CNNs networks were selected for the whole hallucination architecture as the baselines. Garcia {\emph{et al.}} designed a multimodal KD framework that utilizes both depth and RGB videos to learn representations. During testing, images are simultaneously processed by both the RGB and hallucination networks to enhance detection performance. This allows the proposed method to transfer information typically derived from depth training data to a network capable of extracting similar information from RGB data \cite{garcia2018modality}. The entire framework utilizes the ResNet-50-based model as the baseline architecture for each stream block within the framework to mitigate modality discrepancies. Similarly, another work learned sound presentations by transferring knowledge from video to sound modality \cite{aytar2016soundnet}. Similar CNN architectures were employed for both the sound and video recognition networks to promote the extraction of similar features, enhancing knowledge transfer between heterogeneous modalities. Andonian {\emph{et al.}} adopted progressive self-distillation and soft image-text alignments to more efficiently learn robust representations from noisy data for cross-modal contrastive learning.  The framework distilled its own knowledge to dynamically generate soft-alignment targets for a subset of images and captions in every mini-batch, which were subsequently used to update its parameters \cite{andonian2022robust}. 

More recently, Thoker \emph{et al.} proposed a multimodal KD framework for the HAR task. They used RGB videos to train the teacher CNNs network and then trained two student CNNs networks were trained using mutual learning to improve the performance \cite{thoker2019cross}. Furthermore, Quan \emph{et al.} proposed a Semantic-aware Multimodal Transformer Fusion Decoupled Knowledge Distillation (SMTDKD) method, enhancing video data recognition by facilitating information interaction not only between different wearable sensor data but also between visual sensor data and wearable sensor data. To address modality discrepancies and promote the extraction of similar semantic features, graph cross-view attention maps were constructed across different convolutional layers to improve the feature alignment process for HAR problem \cite{quan2023smtdkd}. More recently, Ni {\emph{et al.}} present the first multimodal KD approach for the WSHAR  problem \cite{ni2022cross}. In this study, to enable visual recognition of time series data from IMU sensors, one-dimensional action data from wearable sensors were transformed into visual representations to preserve the local temporal relation. Subsequently, an effective transfer of complementary information from the video domain to the sensor domain was achieved using the same visual VGG-16 models. In order to eliminate the privacy concern from video streams, they further adopted skeleton sequence modality as the teacher model to distill knowledge to time-series modality for accurate WSHAR problem \cite{ni2022progressive}. Specifically, both the teacher and student models employed identical GNN architectures to address the feature alignment problem.

\begin{figure*}[t]
  \centering
\includegraphics[width=\textwidth]{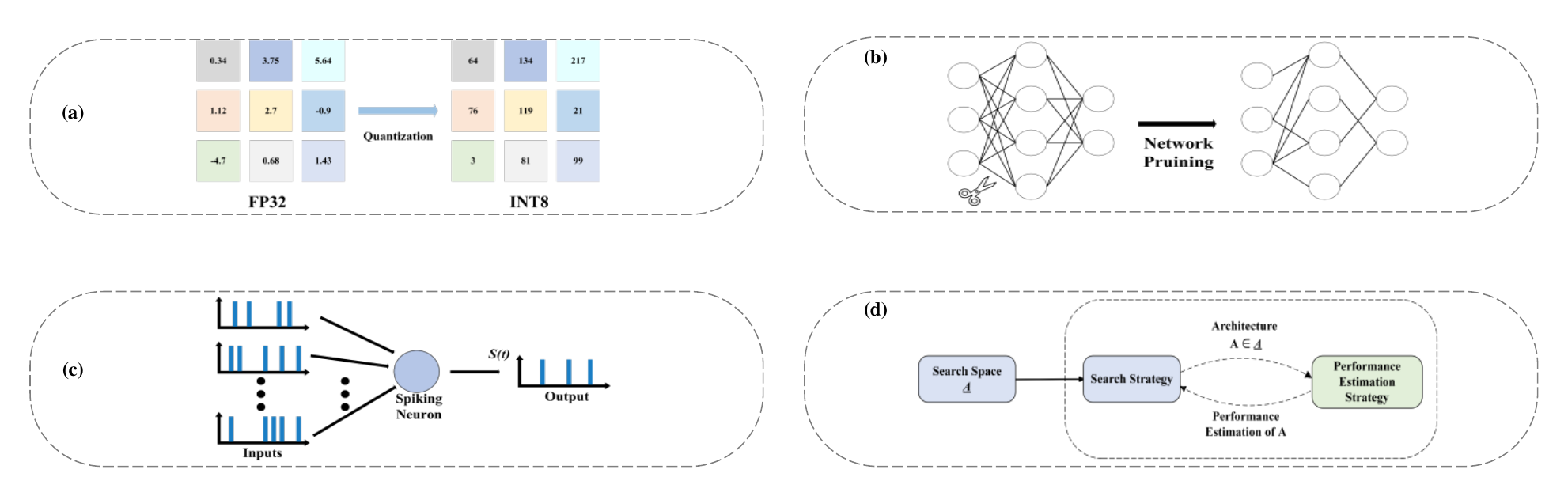}
  \caption{Advanced approaches for model deployment problem. (a) Weight quantization approach \cite{zhu2023survey}. (b) Network pruning method \cite{zhu2023survey}. (c) Spiking neural network architecture \cite{eshraghian2023training}. (d) Neural architecture search framework \cite{elsken2019neural}.}
  \label{figure6}
  \vspace{-0.3 cm}
\end{figure*}

\subsection{Model Deployment Challenge}

While deep learning approaches have demonstrated promising performance in recognizing human activities through IMUs data, they often require substantial resources. Furthermore, wearable technology typically possesses constrained computational capabilities, which obstructs the extensive implementation of models. For instance, a prior study demonstrated that a smartphone (LG Nexus 5X, 1.8 GHz, Hexa-core processor with 2G of RAM) can only sustain a long short-term memory (LSTM) model comprising an input layer, two hidden layers, and an output layer \cite{mauldin2020ensemble}. Therefore, addressing the challenge of significant computational expense is crucial for enabling instantaneous and dependable recognition of human activities on mobile devices using advanced multimodal techniques.

\subsubsection{Multimodal Model Compression}

At present, there exist numerous recognized methods for the purpose of model compression. These include, but are not limited to, techniques such as pruning, knowledge distillation, quantization, and low-rank factorization \cite{zhu2023survey} as shown in Figure \ref{figure6}.a and \ref{figure6}.b. These approaches aim to convert large, resource-intensive models into smaller versions suitable for storage on resource-limited mobile devices. Nonetheless, present compression techniques are mainly constructed on the basis of unimodal networks or particular model architectures, making it challenging to broaden these methods to a variety of multimodal learning methods. A straightforward approach involves training the large-scale model first and then applying post-compression techniques to reduce its size \cite{nooruddin2023multi}. For instance, Nooruddin \emph{et al.} \cite{nooruddin2023multi} first proposed a two-stream multi-resolution fusion architecture for HAR problem. Two quantization approaches, such as post-training quantization and quantization aware training were further introduced to optimize these models for deployment in edge devices. 

Meanwhile, attention mechanism was adopted for lightweight model design process \cite{zhou2022tinyhar,gao2023mmtsa,cai2023acf}. For example, Zhou \emph{et al.} present a lightweight model which included a cross-chanel interaction Transformer encoder and global temporal extraction layer for HAR task \cite{zhou2022tinyhar}. The experiment results showed that the proposed model even surpassed the optimized  DeepConvLSTM \cite{bock2021improving} with reduced model size by more than 93\% on several datasets. Gao \emph{et al.} present a multimodal temporal segment attention network for HAR problem using RGB video and IMU data. This network was tested on a Raspberry Pi 4B, which was equipped with a 64-bit 1.5GHz quad-core CPU and 8GB of RAM. The results of the experiment showed that the network had significantly reduced FLOPs and latency, making it more suitable for edge deployment \cite{gao2023mmtsa}. Cai \emph{et al.} present an adaptive compression framework to address the computational resource challenges which enable input-dependent runtime compression locally on resource-constrained embedded devices \cite{cai2023acf}. Specifically, They propose an offline model transformation module to upgrade the static network with two kinds of dynamic components to support online structural adjustment. A lightweight policy network was designed to generate multi-granularity and data-dependent compression strategies for different model parts.

Another approach is to use specialized toolsets for simplifying complicated models, such as TensorFlow Lite, Caffe2, Pytorch Mobile and TensorRT \cite{luo2020comparison,bursa2023building,mazzia2022action,kang2023efficient}. For instance, TensorFlow Lite is a dedicated suite of tools designed for use on mobile and IoT devices. It provides post-training quantization methods which reduced the model weights using fewer bites instead of full floating-point numbers \cite{luo2020comparison}. Consequently, Bursa \emph{et al.} conducted performance evaluations on a variety of model architectures that were converted using TensorFlow Lite for HAR problem. The experiment results revealed that the application of quantization methods in TensorFlow Lite led to a substantial reduction in model sizes. Importantly, this size reduction did not compromise the accuracy of the models, demonstrating the effectiveness of TensorFlow Lite’s optimization techniques \cite{bursa2023building}.  Mazzia \emph{et al.} proposed a self-attentional architecture that leverages pose representations over small temporal windows. Once converted using TensorFlow Lite, the method provides a low-latency solution that ensures both accuracy and efficiency in real-time performance \cite{mazzia2022action}. In \cite{kang2023efficient}, to demonstrate real-world deployment and applications, the authors utilized TensorRT, an SDK for high-performance DL inference, to convert the trained GNN HAR models based on PyTorch into a TensorRT model for Jetson AGX Xavier and into CoreML for iPhone XR. Experimental results on latency showed that models compressed by TensorRT could reduce latency by approximately 10\% and 50\% on AGX and iPhone XR, respectively.

\subsubsection{Advanced Lightweight Model}

While contemporary DL methodologies have made great progress, they continue to face substantial energy requirement challenges which associated with the training and inference processes. As a result, there has been a surge of interest in developing low-power techniques \cite{hsieh2022ultra}, such as brain-inspired spiking neural networks (SNNs) as shown in Figure \ref{figure6}.c, which leads to high energy efficiency. This is largely due to their event-driven nature, where computations are performed only when events (or spikes) occur \cite{eshraghian2023training}. Currently, there are several studies using SNNs for multimodal learning tasks\cite{sengupta2018integrating,liu2022event,wang2023sstformer,wang2023event,guo2023Transformer}. For instance, Sengupta \emph{et al.} proposed an SNN to fuse temporal, spatial, and orientation data for multimodal brain data modeling. The proposed framework was assessed qualitatively and quantitatively using artificially created data to understand its behavior and its capacity to incorporate spatial, temporal, and orientation data \cite{sengupta2018integrating}. Liu \emph{et al.} integrated SNNs with attention mechanism to fuse visual and auditory data. This attention feature evaluates the importance of each modality and subsequently distributes weights between the two modalities \cite{liu2022event}. Wang \emph{et al.} adopted SNNs to merge RGB frames and event streams concurrently for a pattern recognition problem \cite{wang2023sstformer}. Wang \emph{et al.} designed an event-enhanced multimodal spiking actor network based on deep reinforcement learning. This network combined data from both the Laser and event camera to extract and fuse more effective information \cite{wang2023event}. Guo \emph{et al.} proposed a framework that combined SNNs with Transformer architectures for a multimodal audiovisual classification problem. This framework achieved commendable accuracy in multimodal classification tasks while maintaining low energy usage, positioning it as an efficient and effective solution for such classification tasks \cite{guo2023Transformer}. 

Meanwhile, there are a couple of studies utilizing SNNs for HAR problem \cite{fra2022human,khan2023privacy,li2024snnauth}. For example, Fra \emph{et al.} compared optimized classifiers based on traditional DL architectures and demonstrated the efficiency of SNNs in processing time-dependent signals for HAR problems by not only yielding high performances at a low energy cost \cite{fra2022human}. Specifically, spiking CNNs exhibited the lowest energy consumption at 5.49 $\mu$J, nearly three orders of magnitude less than the corresponding CNN models on Intel’s Loihi. Khan \emph{et al.} integrated SNNs with LSTM networks to achieve energy efficiency and preserve privacy in HAR problem \cite{khan2023privacy}.  Specifically, the proposed spiking LSTM showed a significant improvement in energy efficiency of 32.30\%, compared to simple LSTM. Li \emph{et al.} further present a continuous authentication system which employed SNNs to analyze biometric behavioral patterns recorded by smartphone sensors \cite{li2024snnauth}. To achieve the conversion of ANNs to SNNs, weights and activations were mapped to generate suitable spike neuron models and synaptic connections with higher accuracy performance.

\subsubsection{Neural Architecture Search}

Neural architecture search (NAS) is a process designed to automatically identify the most effective neural network structures that deliver optimal performance while using minimal computational resources \cite{elsken2019neural,
ren2021comprehensive,hsieh2021neural} as shown in Figure \ref{figure6}.d. Currently, several studies utilize NAS for multimodal tasks \cite{perez2019mfas,yu2020deep,xu2021mufasa,shi2021efficient,si2023violence}. For instance, Perez \emph{et al.} introduced a generic search space that encompassed a wide range of potential fusion architectures. A sequential model-based exploration method was designed to find the optimal architecture in the proposed search space. The experiment results demonstrated the benefits of framing multimodal fusion as a problem of neural architecture search \cite{perez2019mfas}. Yu \emph{et al.} devised a generalized NAS framework across several multimodal learning tasks, including visual question answering, image-text matching, and visual grounding \cite{yu2020deep}. The framework was based on a deep encoder-decoder, where each block of the encoder or decoder corresponds to an operation selected from a pre-established operation pool. By employing a gradient-based NAS algorithm, they efficiently learned optimal architectures for different tasks. Xu \emph{et al.} proposed a NAS algorithm to simultaneously search across multimodal fusion strategies and modality-specific architectures for electronic health records diagnosis code prediction \cite{xu2021mufasa}. Shi \emph{et al.} proposed an efficient automatic speech recognition method that benefits from the natural advantage of differentiable NAS in reducing computational overheads. This differentiable architecture search method was fused with Conformer blocks to form a complete search space \cite{shi2021efficient}. Si \emph{et al.} designed a multimodal fusion architecture search framework to automatically design promising multimodal fusion architectures for violence detection tasks \cite{si2023violence}. Specifically, multilayer neural networks based on attention mechanisms are meticulously constructed to grasp intricate spatio-temporal relationships and extract comprehensive multimodal representation. 

More recently, NAS methods have been applied in HAR domain \cite{wang2021harnas,lim2023efficient}. For instance, Wang \emph{et al.} \cite{wang2021harnas} adopted a multi-objective NAS method to solve the tradeoff problem between high efficiency and high performance. This framework was extended to a tri-objective task where the search targets were based on the weighted F1 score, the number of FLOPs, and its memory use. Lim  \emph{et al.} proposed a mobile HAR NAS based on a differentiable neural architecture search for automatic design of the architecture of a HAR model for a mobile device \cite{lim2023efficient}. Experiments were also conducted with the Galaxy A31 and Galaxy S10 smartphones as target devices. the latency of the A31-optimized model was, on average, 2\% faster than that of the S10-optimized model on the A31 device.

\section{Future Research Direction}

Despite the efforts devoted to these above-mentioned challenges, some of them are still not fully explored, such as heterogeneous feature alignment, lightweight model deployment, and so on. While existing research may not offer complete and dependable solutions to these challenges, they do provide a solid base and valuable insights for future work. Additionally, there are several other challenges that have been scarcely investigated before and require immediate exploration. In the following section, we highlight several pivotal research directions that urgently need exploration. It is our hope that the challenges identified in this study can serve as catalysts for these future explorations.

\subsection{Future Activity Prediction}

Future activity prediction can be considered as an extension version of current HAR problem. Unlike activity recognition, which identifies current actions, the predictive system can anticipate user behaviors beforehand. This system plays a crucial role in understanding human intentions, thereby finding applications in smart services, crime detection, and driver behavior prediction. In some common behavior tasks, the activities are usually in a certain order. Therefore, modeling the temporal dependencies across activities is beneficial to predict future predictions. Cross-modal knowledge distillation framework \cite{wang2019progressive} is suitable for such tasks. But for long-span activities captured from partial video, KD cannot achieve such long dependencies due to the limited context information. In this case, adversarial KD-based approach based on generative network can assist to solve the early action prediction task \cite{zheng2023egocentric}. Moreover, LLMs have shown considerable potential in identifying patterns, predicting future events, and detecting anomalous behavior across diverse domains \cite{su2024large,zhang2024mm}. Consequently, exploring the potential of leveraging LLMs for future WSHAR prediction emerges as a promising avenue for research.

\subsection{Identifying Unknown Activities}

Discovering unprecedented actions that remain unobserved by the models poses a substantial obstacle in HAR domain. Hence, it becomes imperative to examine the capacity of models to adjust to dynamic environments and prevent the disastrous loss of previously acquired knowledge. In fact, an effective model should have the ability to acquire new insights in an online manner and execute accurate discernment in the absence of ground truth. A promising way to enable models to continuously adapt to dynamic input data, is continue learning \cite{gammulle2023continuous}. However, how to build models with the ability to perpetually adapt to multi-model data still an under-explored problem. Moreover, LLMs have shown significant achievements to directly comprehend visual signals. For example, LLMs can fundamentally considers images as linguistic entities, translating them into discrete words from the LLM's vocabulary \cite{zhu2024beyond}. More recently, the semantic space of LLMs has demonstrated to be able to guide time series embeddings by maximizing the cosine similarity in the joint space \cite{pan2024textbf}. Therefore, exploring LLMs fully for identifying unseen activities can be another promising research area for HAR improvements.

\subsection{New Foundation Models}

Currently, Transformers have dominated the HAR domain \cite{sun2022human}. However, one of the disadvantages of Transformer model is the computational inefficiency on long sequences data. As a result, a new model called Mamba has been proposed to solve this problem \cite{gu2023mamba}. In this study, a selection mechanism to structured state space models is designed to engage in context-dependent reasoning while maintaining linear scalability in sequence length. Mamba and its variants have showcased the extensive applicability of selective state space models in modalities requiring extensive context, such as audio, image, and video \cite{ma2024u,liang2024pointmamba,li2024videomamba}. Meanwhile, RWKV \cite{peng2023rwkv} integrated the efficient parallelizable training of Transformers with the effective inference capabilities of RNNs, maintaining constant computational and memory complexity during inference. RWKV and its variants have exhibited comparable performance as well as promising latency and memory utilization efficiency \cite{duan2024vision,an2023exploring,hou2024rwkv}. Further research into the application
of these foundation models in the WSHAR  domain could potentially lead to the development of more robust models.

\subsection{Unified Multimodal Systems}

Existing papers that apply multimodal learning approaches for wearable HAR problem mainly concentrates on the fusion of diverse inputs from different modalities and a single task at a time, such as forecasting and classification. However, these studies do not facilitate the simultaneous analysis of multimodal and multitask scenarios. In the domains of CV, NLP and audio, models such as Unified-io \cite{lu2022unified}, AnyGPT \cite{zhan2024anygpt}, and UniAudio \cite{yang2023uniaudio} have integrated  multiple input modalities to support the execution of multiple tasks within a singular Transformer-based architecture. Omni-Dimensional INstance segmentation (ODIN) method was proposed to segment and label both 2D RGB images and 3D point clouds \cite{jain2024odin}. This approach utilizes a Transformer architecture that alternates between 2D within-view and 3D cross-view information fusion. Similarly, a one-for-all model, called UniST \cite{yuan2024unist}, was proposed to solve for urban spatio-temporal prediction problem. More recently, UniTS \cite{gao2024units} showed its capable of handling various tasks such as forecasting and anomaly detection through a universal task specification. Further research into the application of multimodal and multitask analysis in the HAR domain could potentially lead to the development of more potent time series foundation models.

\subsection{Concurrent Activity Segmentation}

Human activities naturally exhibit a hierarchical structure, as seen in daily routines that include a variety of tasks such as washing, grooming, and eating. These tasks can be further divided into specific actions like washing dishes or washing hands. However, compiling a large-scale dataset of everyday activities with detailed annotations is a challenging task, mainly due to the time-consuming requirement for manual annotations. Furthermore, distinguishing between activities with comparable performance trends poses a challenge for online models in wearable HAR domain. In CV and audio domain, models such as SAM \cite{kirillov2023segment} and AV-SAM \cite{mo2023av} achieved amazing segmentation performance and how to leverage such method to wearable HAR area is a promising direction for future research. Furthermore, LLMs can be employed to generate descriptions of both overarching routines and their associated detailed actions, potentially serving as annotations for the synthesized virtual IMU data. This advanced hierarchical dataset can be further utilized to train a model capable of identifying and segmenting various activity levels, thereby enhancing the performance of WSHAR systems.

\subsection{Personalization and Privacy}

The majority of existing research on wearable HAR  and time series analysis usually focuses on a global model for all users. However, the development of personalized models for individual users, derived from the global model, could potentially offer additional advantages and adaptability. This approach could lead to more tailored solutions that better meet the unique needs of each user. Moreover, privacy is indeed a crucial factor, particularly as a significant amount of time series data is gathered in private contexts for purposes such as clinical applications or smart home technologies. As a result, federated learning is applied to build personalized model for wearable HAR problem using single modality \cite{xiao2021federated}. However, the task of utilizing federated learning frameworks in the context of multimodal HAR remains a complex and less explored area. Meanwhile, style transfer in CV area involves generating a new image by combining the content of one image with the style of another image \cite{gatys2015neural}. The goal of style transfer is to create an image that preserves the content of the original image while applying the visual style of another image. As a result, style transfer approach can be a valuable direction for personalized sample regeneration in time series domain. In fact, advancing research into multimodal personalized model and user privacy preservation would broaden the scope and utility of multimodal HAR problem.

\section{Conclusion}

We present the first survey that systematically analyzes the WSHAR field from the perspective of multimodal learning approaches. Initially, we discuss
recent advancements in sensor modalities and the latest deep learning approaches for HAR. Then, we explore recent techniques used in present multimodal systems for WSHAR, covering both inter-multimodal systems utilizing sensor modalities from visual and non-visual systems, and intra-multimodal systems using modalities from non-visual systems only. In addition to providing a comprehensive summary of existing multimodal datasets for WSHAR, we also discuss the accomplishments of multimodal approaches in addressing some of the challenges in the WSHAR field. By connecting the existing multimodal achievements from other tasks, such as CV and NLP domains, we have laid the groundwork for discussions on the existing challenges and potential future directions. This paper concludes with final remarks that encapsulate the essence of our findings and discussions. We hope that our work will inspire further research in this exciting and rapidly evolving research field.

\vfill


\vfill\pagebreak
\bibliographystyle{IEEEtran}
\bibliography{sample-base}
\end{document}